\documentstyle[12pt]{article}
\begin{document}
\def\theequation{\arabic{section}.\arabic{equation}}
\newcommand{\be}{\begin{equation}}
\newcommand{\ee}{\end{equation}}
\begin{titlepage}
\title{On the rotation of polarization by a gravitational
lens}
\author{Valerio Faraoni
\\{\small \it Department of Physics and Astronomy,
University of Victoria}\\
{\small \it P.O. Box 3055, Victoria, B.C. V8W~3P6 (Canada)}}
\date{}
\maketitle
\thispagestyle{empty}
\begin{abstract}
It is proved that the field of a gravitational lens induces no rotation
in the polarization vector of electromagnetic radiation, in agreement
with the previous literature, but with a different approach. The
result is generalized to the case of less conventional gravitational
lenses (static cosmic strings and gravitational waves).
\end{abstract}
\vspace*{5truecm}
\begin{center}
To appear in {\em Astronomy and Astrophysics}
\end{center}
\end{titlepage}
\clearpage
\setcounter{page}{1}
\section{Introduction}
The problem if propagation of polarized electromagnetic radiation through
the gravitational field of a cosmic object is accompanied by a rotation of
the plane of polarization of the radiation, has been considered
recently (Dyer \& Shaver 1992, hereafter DS). If such an effect would exist,
it could be relevant for the
analysis of photons coming from a gravitationally lensed source, or
for the microwave background photons propagating in a inhomogeneous
universe. It has been proposed (Kronberg et al. 1991) that polarization
data from gravitationally lensed radio sources can be used to probe
the mass distribution in the lens. In the paper by Kronberg et al. (1991),
it is assumed that the polarization vector of the electromagnetic
radiation is not rotated by the gravitational lens. This assumption
has been discussed in a later paper by DS, who have proved it for most
astrophysically
interesting lenses, mainly using symmetry arguments in a relativistic
context. Our purpose is to compare the result by
DS with an alternative approach, adopting a
post-Newtonian metric to describe spacetime, and a Newtonian potential
to describe the gravitational field of the lens, which are more
conventional tools in gravitational lens theory. Calculations are
performed in Sec.~2, and the result by DS is confirmed.
In Sec.~3, we consider the case of more exotic gravitational lenses
which have been treated in the literature, namely cosmic strings
and gravitational waves. It is shown that the negative result of
Sec.~2 holds also for these nonconventional lenses. A symmetry
argument analogous to those in DS' paper is used for the case of a
static, straight, infinite cosmic string, while for gravitational
waves the calculation is closely analogous to that for ordinary
gravitational lenses.
\section{Gravitational lenses and polarized radiation}
Let us consider a radio source which is lensed by a gravitational
lens described by the Newtonian potential $ \phi $; we assume that
\begin{enumerate}
\item the spacetime is described by the post-Newtonian metric\footnote{The
metric signature is $+2$. Greek indices run from $0$ to $3$, Latin
indices run from $ 1$ to $ 3 $. We use units in which $ G=c=1 $, and
perform computations to first order in $ \phi $ and its
derivatives.}\setcounter{equation}{0}
\be
\label{1}
g_{\mu \nu}=\eta_{\mu\nu}+h_{\mu\nu} \; ,
\ee
where $ \eta_{\mu\nu} $ are the components of the Minkowski metric in
an asymptotically Cartesian coordinate system $\left\{ x^{\mu}
\right\}=\left\{ t,x,y,z \right\} $, and
\begin{eqnarray}
\label{2}
& & h_{00}=-2\phi \; , \\
\label{3}
& & h_{ij}=-2\phi  \,\delta_{ij} \; ,
\end{eqnarray}
are small perturbations;
\item geometric optics holds;
\item light rays suffer only small deflections;
\item the lens is bounded and stationary, i.e.
\be
\label{4}
\phi (x,y,z) \rightarrow 0 \;,\;\;\;\;\;
\partial_i \phi \rightarrow 0 \;\;\;\;\;\;\;
\mbox{as}\;\;\;\;\;\;\;r\equiv \left( x^2+y^2+z^2 \right)^{1/2}
\rightarrow +\infty \; ;
\ee
\be
\label{5}
\frac{\partial \phi}{\partial t} \simeq 0 \;.
\ee
\end{enumerate}
In the present calculations, we assume that the observer is an inertial
observer of the Minkowski background, with four velocity components $
u^{\mu}=\delta^{0\mu} $ in the $ \left\{ x^{\alpha} \right\} $
coordinate system. As customary in gravitational lens studies, cosmology
can be fit into the
model at a later stage, by using angular diameter distances in a
Friedmann-Lemaitre-Robertson-Walker universe to measure the distances
between observer, lens, and source. We will actually need to go beyond
the
geometric optics approximation, in order to describe polarization of
the electromagnetic radiation.

An electromagnetic wave in a curved spacetime can be described by the
Maxwell four-potential
\be
\label{14}
A_{\mu}=\hat{A}_{\mu} ( x^{\alpha}) \, e^{i \, \omega S( x^{\alpha})} \;,
\ee
where $\hat{A}^{\mu} $ is a complex vector field, $ S $ is a real
function (the eikonal), and $ \omega $ is the frequency of the
electromagnetic wave. $A^{\mu} $ and $ S^{\mu}\equiv S^{,\mu} $
satisfy (Stephani 1982)
\begin{eqnarray}
\label{15}
& & S_{\mu}S^{\mu}=0 \; , \\
\label{16}
& & A_{\mu}S^{\mu}=0 \; ,  \\
\label{17}
& & S^{\nu}\nabla_{\nu}S^{\mu}=0 \; .
\end{eqnarray}
The complex vector $ \hat{A}^{\mu} $ can be decomposed into its complex
magnitude $ a $ and the real polarization vector $ P^{\mu} $
\be
\label{18}
\hat{A}^{\mu}=aP^{\mu} \; ,
\ee
where $ a $ and $ P^{\mu} $ obey the equations (Stephani 1982)
\begin{eqnarray}
\label{19}
& & \frac{1}{a}\, \frac{da}{d\lambda}=-\, \theta \; , \\
& & \frac{dP^{\mu}}{d\lambda}=\frac{1}{2} \left(
\frac{P^{\nu} \partial_{\nu}a}{a} +\nabla_{\nu}P^{\nu} \right) S^{\mu}  \;.
\label{20}
\end{eqnarray}
Here, $ \lambda $ is an affine parameter along the null geodesics, and
$ \theta \equiv \nabla_{\alpha}S^{\alpha}/2 $ is the expansion of the
congruence of null geodesics around a fiducial ray, which obeys the
well known Raychaudhuri's equation. We can write
\begin{eqnarray}
\label{21}
& & a=a^{(0)}+\delta a \; , \\
& & P^{\mu}=P^{(0) \mu}+\delta P^{\mu} \; ,
\label{22}
\end{eqnarray}
where $ \delta a $ and $ \delta P^{\mu} $ are small perturbations to
the flat space quantities $ a^{(0)} $ and $ P^{(0) \mu} $.

We choose, as a solution to Eqs. (\ref{19}) and (\ref{20}) in
the unperturbed case (for which $\theta^{(0)}=1/\lambda $),
\begin{eqnarray}
\label{23}
& & a^{(0)}=\frac{A}{\lambda} \; , \\
\label{24}
& & P^{(0) \mu}=(0,1,0,0) \; ,
\end{eqnarray}
where $ A $ is a complex constant, the unperturbed photons are
polarized along the $ x$-axis, and the affine parameter $ \lambda=t $ is
measured from the source position. We assume now that the optic axis of the
lens coincides with the $z$-axis, and consider a bundle of null rays
such that the corresponding photons have unperturbed paths parallel to it,
and are only slightly deflected by the gravitational lens. The four-vector
tangent to a perturbed path is
\be
\label{6}
S^{\mu}=S^{(0) \mu}+\delta S^{\mu}=(1+\delta S^0,\delta S^1, \delta
S^2, 1+\delta S^3 ) \; ,
\ee
where $ S^{(0) \mu} \equiv (1,0,0,1) $ is the tangent to the unperturbed
path. The equation of null geodesics
\be
\label{8}
\frac{dS^{\mu}}{d \lambda}+\Gamma^{\mu}_{\rho \sigma} S^{\rho}
S^{\sigma} =0 \; ,
\ee
and
\be
\label{9}
\Gamma^{\mu}_{\rho \sigma}=\frac{1}{2} \,\eta^{\mu \alpha}
\left( h_{\alpha \rho ,\sigma}+h_{\alpha \sigma ,\rho}-
h_{\rho \sigma ,\alpha} \right) \; ,
\ee
give the following expression for the deflection $\delta S^{\mu} $
\be
\label{10}
\delta S^{\mu}=-\int_{S}^{O} d\lambda \left( {h^{\mu}}_{\rho ,\sigma}
-\frac{1}{2} \, {h_{\rho \sigma}}^{,\mu} \right) S^{(0) \rho}S^{(0) \sigma}
+O(2) \; ,
\ee
where the integral in the right hand side is computed along the perturbed
photon path from the source to the observer. Equation (\ref{10}) gives,
to first order,
\begin{eqnarray}
\label{11}
& & \delta S^0=0 \; , \\
\label{12}
& & \delta S^i=-2\int_{S}^{O} dz \, \partial_{i} \phi +O(2) \; ,\\
& & \delta S^3=0
\label{13}
\end{eqnarray}
($ i=1,2 $), where the integral in the right hand side of Eq.~(\ref{12})
is now computed along the {\em unperturbed} photon path from the source to
the observer (performing the integration along the actual path only adds
a second order contribution). The gravitational lens
induces no frequency shifts and no deflections in the $ z $-direction, to
first order, as can be seen from Eqs.~(\ref{11}) and (\ref{13}), and
is well known.

The perturbations $ \delta a $ and $ \delta P^{\mu} $ satisfy
\begin{eqnarray}
\label{26}
& & \frac{1}{a^{(0)}} \,\frac{d( \delta a)}{d\lambda}+
\frac{1}{\lambda}\,\frac{\delta a}{a^{(0)}}+\delta \theta
=0 \; , \\
& & \frac{d( \delta P^{\mu})}{d\lambda}=\frac{1}{2} \left[
 \frac{P^{(0) \nu}\partial_{\nu}( \delta a)}{a^{(0)}}+
\nabla_{\nu}P^{\nu} \right] S^{\mu}  \; ,
\label{27}
\end{eqnarray}
with $ \delta \theta \equiv \theta -\theta^{(0)} $ given by the
Raychaudhuri's equation.
For our purposes, it is not necessary to solve the coupled system of
Eqs.~(\ref{26}), (\ref{27}), but it is sufficient to note that, using
Eqs.~(\ref{5}), (\ref{23}), (\ref{13}), and
\be
\label{28}
\sqrt{|g|}=1-2\phi +O(2) \; ,
\ee
Eq.~(\ref{27}) reduces to
\be
\frac{d( \delta P^{\alpha})}{d \lambda}=\frac{1}{2} \left[
\frac{\partial_{x}( \delta a)}{a^{(0)}}+\partial_{\mu}( \delta P^{\mu})
-2\,\partial_{x}\phi \right] \left( \delta^{0\alpha}+\delta^{3\alpha}
\right) \; .
\label{30}
\ee
We have, from Eq.~(\ref{30}),
\be
\label{31}
\frac{d( \delta P^1)}{d\lambda}=\frac{d( \delta P^2)}{d\lambda}=0
\ee
and, since $\delta P^{\mu} \rightarrow 0 $ as $ r\rightarrow + \infty
$ (at least approaching the source of radiation), we get
\be
\label{32}
\delta P^1=\delta P^2=0 \; .
\ee
Since $ P^{\mu} $ is a purely spatial vector, $ \delta P^0=0 $. Moreover,
Eq.~(\ref{30}) gives
\be
\label{33}
\frac{d( \delta P^0)}{d\lambda}=\frac{d( \delta P^3)}{d\lambda} \; ,
\ee
and we conclude that $ \delta P^3 =\delta P^0=0 $ as well. Thus, to first
order, the gravitational lens induces no changes in the polarization
vector of electromagnetic radiation:
\be
\label{34}
\delta P^{\mu}=0 \; ,
\ee
in agreement with the result by DS.
\section{Discussion and conclusion}
The action of a gravitational lens on the plane of polarization of
electromagnetic radiation from a distant source has been found to be
negligible to first order, according to the result by DS. We describe the
gravitational lens with a Newtonian potential, as customary in
the conventional gravitational lens theory
(Bourassa et al. 1973; Bourassa \& Kantowski 1975; Schneider 1985;
Blandford \& Narayan 1986; Blandford
\& Kochanek 1988). This approach is
alternative to that of DS, but it is obviously
limited to situations in which the assumptions needed to apply the standard
scalar or vector formalism (based on the use of a post-Newtonian
metric) hold. In the case of rotating or, more generally,
nonstationary
gravitational lenses, these formalisms cannot be employed, or their
use requires a particular care (Faraoni 1991). However, it seems that most
objects of astrophysical interest whose action as gravitational
lenses is likely to be observed, behave as static, and can therefore
be described by the vector or scalar formalism. This class of objects
include conventional lenses like galaxies, galaxy clusters, microlenses
like stars, Jupiters, binary systems, and even compact objects like
neutron stars or black holes, provided that lensing takes place in a
region where the field is sufficiently weak (a situation much more
more likely than lensing in a strong field region). The velocities of
galaxies in clusters, of stars in galaxies, and the rotational
velocities of double galaxies, or binary systems of stars, planets,
or collapsed objects (provided that the system is not an extremely close
binary), are not high enough to make the lens nonstationary
(Faraoni 1991). The case of
lensing by an extremely rotating lens should be considered as highly
unlikely in the real world, as pointed out in DS. A Schwarzschild lens,
though static, cannot be described by the scalar or vector formalism of
gravitational lensing theory, except for the weak field regime. In the
strong field region, the symmetry arguments of DS are particularly
convenient.

In principle, one can also conceive of more exotic lenses. The
gravitational field due to density perturbations which are collapsing
to form structures in the early universe is by far too weak to affect
appreciably the polarization of the microwave background photons
propagating close to them. Another possibility is represented by
lensing cosmic strings (Vilenkin 1984; Hogan \& Narayan 1984;
Paczynski 1986; Birkinshaw 1989). The spacetime
around a straight, static, infinite string can be described by the
metric\setcounter{equation}{0}
\be
\label{string}
ds^2=-dt^2+dr^2+\left( 1-8\mu \right) r^2 d\varphi^2+dz^2 \; ,
\ee
where $ \mu $ is the linear mass density of the string. This metric
can be brought to the Minkowskian form by introducing
the coordinate $ \varphi '\equiv (1-4\mu ) \, \varphi $, which varies in
the range $ \left[ 0, \left(1-4\mu \right) 2\pi \right] $. Though
spacetime is locally flat, the existence of a deficit angle $ \delta =
4 \mu $ causes the deflection of a light ray propagating in the $ xy
$ plane, and the formation of a double image of a distant source. By
using symmetry arguments analogous to those of DS,
one concludes that the polarization vector of the electromagnetic
radiation is not affected by the string, since the latter does not
introduce any preferred verse of rotation of the polarization vector
(one can reach
the same conclusion by using Eqs.~(\ref{19}) and (\ref{20}) in the
metric given by Eq.~(\ref{string})). A possible
exception is given by a cosmic string moving at relativistic velocity
(Birkinshaw 1989); in this case the symmetry argument does not hold.
However, no observational evidence has been given for such objects,
and we will not investigate this situation here.

Another possibility considered in the literature is lensing by
gravitational waves, both considered
as lens components superposed to the caustic structure of an ordinary
gravitational lens (McBreen \& Metcalfe 1988; Allen 1989, 1990; Kovner
1990), or as full lenses (Wheeler 1960; Zipoy 1966; Zipoy \& Bertotti
1968; Kaufmann 1970; Bertotti 1971; Dautcourt 1974; Bertotti and
Catenacci 1975; Burke 1975; Linder 1986, 1988; Braginsky et al. 1990;
Faraoni 1991, 1992). As fas as these
situations are concerned, we remind the reader that the shear induced
by a gravitational wave creates an
anisotropy in the plane orthogonal to direction of propagation of the
wave. This results in a polarization of the order of the metric
perturbations describing the gravitational wave; for astrophysically
generated gravitational waves, this effect is too small to be
detectable with present techniques.
In the case of long wavelength gravitational waves of cosmological
origin (which cannot be treated in the context of the thin lens
approximation), the
effect on the polarization of microwave background photons has been
taken into account, and has been used to set upper limits on the
cosmological density of such waves (Matzner 1988). However, this
effect is like the photon scattering in an anisotropic medium, and is
quite different from the rotation of the plane of polarization. In fact,
assuming that the spacetime metric is given by
\be
g_{\mu\nu}=\eta_{\mu\nu}+\gamma_{\mu\nu} \; ,
\ee
where $ \gamma_{\mu\nu} $ are small perturbations describing the
gravitational waves, one finds, with calculations closely parallel to
those in Sec.~2,
\be
\frac{d( \delta P^{\alpha})}{d \lambda}=\frac{1}{2} \left[
\frac{\partial_{x}( \delta a)}{a^{(0)}}+\partial_{\mu}( \delta P^{\mu})
+\frac{1}{2}\,\partial_{x}\gamma \right] \left( \delta^{0\alpha}+
\delta^{3\alpha} \right) \; ,
\label{30onde}
\ee
where
\be
\gamma \equiv {\gamma^{\mu}}_{\mu}=-\gamma_{00}+\gamma_{11}+\gamma_{22}
+\gamma_{33}+O(2) \; .
\ee
Equation (\ref{30onde}) is analogous to Eq.~(\ref{30}). The same arguments
used above lead to the result that $ \delta P^{\mu} =0  $ in this case
as well.

As a conclusion, any common kind of gravitational lens does not induce
rotation of polarization, at least at the level detectable by present
techniques.
\section*{Acknowledgments}
Doctor P. Schneider is acknowledged for helpful comments. The author
would like to thank also the warm hospitality at the University of Victoria,
in particular Professor F. I. Cooperstock. This work was supported by the
``Fondazione Angelo della Riccia'', which is gratefully acknowledged.
{\small \section*{References}
Allen, B. 1989, Phys. Rev. Lett. 63, 2017 \\
Allen, B. 1990, Gen. Relativ. Grav. 22, 1447 \\
Bertotti, B. 1971, in Sachs, R.K. (ed.) General Relativity and
Cosmology. Academic Press, New York, p.347 \\
Bertotti, B., Catenacci, R. 1975, Gen. Relativ. Grav. 6, 329 \\
Birkinshaw, M. 1989, in Moran, J.M., Hewitt, J.N., Lo K.Y. (eds.)
Gravitational Lenses, Proceedings Cambridge, Massachusetts, USA, 1988.
Springer, Berlin, p.~59-64\\
Blandford, R.D., Kochanek, C.S. 1988, in Bachall, J.N., Piran, T.,
Weinberg, S. (eds.) Proceedings of the 13th Jerusalem Winter School on
Dark Matter in the Universe. World Scientific, Singapore, p.~133\\
Blandford, R.D., Narayan, R. 1986, ApJ 310, 568\\
Bourassa, R.R., Kantowski, R. 1975, ApJ 195, 13 \\
Bourassa, R.R., Kantowski, R., Norton, T.D. 1973, ApJ 185, 747 \\
Braginsky, V.B., Kardashev, N.S., Polnarev, A.G., Novikov, I.D. 1990,
Nuovo Cimento 105B, 1141 \\
Burke, W.L. 1975, ApJ 196, 329 \\
Dautcourt, G. 1974, in Longair, M.S. (ed.) IAU Symp.~63, Confrontation
of Cosmological Theories with Observation. Reidel, Dordrecht, p.~299 \\
Dyer C.C. Shaver, E.G. 1992, ApJ 390, L5 (DS) \\
Faraoni, V. 1991, preprint SISSA 161/91/A, to appear in ApJ \\
Faraoni, V. 1992, in Kayser, R., Schramm, T., Refsdal, S. (eds.)
Gravitational Lenses, Proceedings, Hamburg, Germany 1991. Springer, Berlin
(in press) \\
Hogan, C., Narayan, R. 1984, MNRAS 211, 575 \\
Kaufmann, W.J. 1970, Nat 227, 157 \\
Kovner, I. 1990, ApJ 351, 114 \\
Kronberg, P.P., Dyer, C.C., Burbidge, E.M., Junkkarinen, V.T. 1991,
ApJ 367, L1 \\
Linder, E.V. 1986, Phys. Rev. D 34, 1759 \\
Linder, E.V. 1988, ApJ 328, 77 \\
Matzner, R.A. 1988, Univ. of Texas preprint \\
McBreen, B., Metcalfe, L. 1988, Nat 332, 234 \\
Paczynski, B. 1986, Nat 319, 567 \\
Schneider, P. 1985, A\&A 143, 413 \\
Stephani, H. 1982, General Relativity: An Introduction to the Theory of
the Gravitational Field. Cambridge Univ. Press, p.75-77 \\
Vilenkin, A. 1984, ApJ 282, L51 \\
Wheeler, J.A. 1960, in Radicati, L.A. (ed.) Rendiconti della Scuola
Internazionale di Fisica ``Enrico Fermi'', 11th Course of the Varenna
Summer School, 1959,``Interazioni Deboli''. Zanichelli, Bologna, p.67 \\
Winterberg, F. 1968, Nuovo Cimento 53B, 195 \\
Zipoy, D.M. 1966, Phys. Rev. 142, 825 \\
Zipoy, D.M., Bertotti, B. 1968, Nuovo Cimento 56B, 195}
\end{document}